\documentclass[11pt]{article}
\usepackage{setspace}
\usepackage{booktabs} 
\usepackage{url}
\usepackage{graphicx}
\usepackage{tabularx}
\usepackage{verbatim}
\makeatletter
\newcommand{\verbatimfont}[1]{\renewcommand{\verbatim@font}{\ttfamily#1}}
\makeatother

\begin{document}

\title{Modeling Data Lake Metadata with a Data Vault}

\author{
Iuri D. Nogueira\\
       Universit\'{e} de Lyon, Lyon 2\\
       France\\
       iuri.deolindonogueira@univ-lyon2.fr	
\and
Maram Romdhane\\
       Universit\'{e} de Lyon, Lyon 2\\
       France\\
       maram.romdhane@univ-lyon2.fr
\and
J\'{e}r\^{o}me Darmont\\
       Universit\'{e} de Lyon, Lyon 2, ERIC EA 3083\\
       5 avenue Pierre Mend\`{e}s France, F69676 Bron Cedex, France\\
       jerome.darmont@univ-lyon2.fr}

\date{}

\maketitle

~\\
\begin{abstract}
With the rise of big data, business intelligence had to find solutions for managing even greater data volumes and variety than in data warehouses, which proved ill-adapted. Data lakes answer these needs from a storage point of view, but require managing adequate metadata to guarantee an efficient access to data. Starting from a multidimensional metadata model designed for an industrial heritage data lake presenting a lack of schema evolutivity, we propose in this paper to use ensemble modeling, and more precisely a data vault, to address this issue. To illustrate the feasibility of this approach, we instantiate our metadata conceptual model into relational and document-oriented logical and physical models, respectively. We also compare the physical models in terms of metadata storage and query response time.
\end{abstract}
~\\


{\bf Topics:} Information systems, Information storage systems, Cloud based storage
~\\

\section{Introduction} 
\label{intro}

Born with big data in the 2010's, data lakes propose a way to store in their native format voluminous and diversely structured data, in an evolutionary storage place allowing later analyses (reporting, visualization, data mining...) \cite{Dixon10}. The concept of data lake opposes that of data warehousing, which outputs an integrated, highly structured, decision-oriented and subject-oriented database, but has the disadvantage of dividing data into silos \cite{SteinM14}.

Yet, everyone agrees that a data lake must be well designed. Otherwise, it is doomed to quickly become an inoperable data swamp~\cite{AlrehamyW15,Inmon2016}. That is, a data lake must allow querying data (selection/restriction) with good response times, and not only storing data and handling a ``key-value'' access. However, actual solutions are more or less non-existent in the literature and pertain to undisclosed industrial practices.

This is why \cite{Pathirana15} proposed a conceptual metadata model that  allows indexing and efficiently querying an industrial heritage data lake constituted of XML data, texts, floorplans and pictures. This multidimensional model is similar to snowflake models used in data warehouses or datamarts, but only stores metadata, not the document corpus itself.

This metadata model has been instantiated physically in several NoSQL database management systems (DBMSs) to enable scaling. However, such a data warehouse-like schema cannot evolve easily when data sources evolve themselves, or when new sources pop up, whereas this is a crucial point in the management of a data lake.

Consequently, we propose in this paper:
\begin{enumerate}
	\item to replace the multidimensional model of \cite{Pathirana15} by an ensemble model, more precisely a data vault \cite{Linstedt11,Hultgren12,eshetu14,LinstedtO15}, which is a data model allowing easy schema evolutions and has, to the best of our knowledge, never been used in the context of metadata management;
	\item to evaluate, the feasibility, on one hand, and the effectiveness of this model in terms of metadata query response (as an index), on the other hand, because it induces many joints. For this sake, we translate our conceptual metadata vault into different logical (i.e., relational and document-oriented) and physical  (i.e., PostgreSQL\footnote{\url{https://www.postgresql.org}} and MongoDB\footnote{\url{https://www.mongodb.com}}) models, which also helps compare the respective efficiency of the two physical models.
\end{enumerate}

The remainder of this paper is organized as follows. Section~\ref{EDA} is devoted to a state of the art concerning data lakes and metadata management, on one hand, and ensemble modeling and data vaults in particular, on the other hand. Section~\ref{modelemeta} presents our conceptual metadata vault model, as well as its logical and physical translations. Section~\ref{expes} details the experiments we conducted to validate the feasibility of our approach and compare the PostgreSQL and MongoDB physical models in terms of storage and metadata query response time. Finally, Section~\ref{conclusion} concludes this paper and hints at research perspectives.

\section{Related Work} 
\label{EDA}

\subsection{Data Lakes and Metadata} 
\label{datalakes}

Data lakes aim at quickly integrate very large volumes of mixed types data, i.e., from structured to unstructured data \cite{Dixon10,OLeary2014,Inmon2016}. However, they are not only a storage technology, i.e., mostly the Hadoop Distributed File System (HDFS), but offer a new data ecosystem that allows cross-analyzing data on demand \cite{Ganore15}, without needing costly preprocessing tasks as in data warehousing processes. Data are immediately accessible through the schema-on-read (or late binding) approach \cite{Fang2015}, unlike, again, in data warehouses that are periodically refreshed through an Extraction, Transformation and Loading (ETL) phase (schema-on-write) that may be expensive \cite{MiloslavskayaT16}.

In summary, data lakes manage data with variety for on-demand, ad-hoc analyses; while data warehouses manage structured data for recurring, industrialized analyses.
 
Data lakes are typically subdivided into three components \cite{SteinM14,AlrehamyW15}: 
\begin{enumerate}
	\item a storage system: very often HDFS, though NoSQL DBMSs are also an option \cite{Inmon2016,Khine2017};
	\item a metadata system, which we focus on below;
	\item an access and analysis system generally relying on MapReduce or Spark \cite{John2017}.
\end{enumerate}
By contrast, \cite{Inmon2016} proposes an architecture with respect to data types, made of so-called data ponds for raw data, analog (stream) data, application data (i.e., a data warehouse), textual data and archived data, respectively. In such a context, database research on querying multistore systems \cite{Bondiombouy16,Tan2017} looks highly relevant.

Data lakes mostly bear a ``flat'' architecture, where each data element has a unique identifier and a set of characterizing tags \cite{MiloslavskayaT16}. Such tags are actually essential metadata to help comprehend data and access data \cite{AlrehamyW15}, not to mention query effectiveness.

The advantage of data tagging is that new data and new sources can be inserted on the fly. Once data are tagged, they just need to be connected to already stored data. This feature allows users to formulate queries directly, without needing the help of a business intelligence expert. Yet, this feature requires that the metadata system can manage this evolutive data structure.

\subsection{Ensemble Modeling and Data Vaults}
\label{datavaults}

Ensemble modeling is an approach dating back from the early 2010's, which aims to renormalize data warehouses to bring them closer to the business concepts they model; and to allow better evolutivity, both in terms of data and schema \cite{anchorVSvault13}.

The two prominent approaches in ensemble modeling are data vaults \cite{Linstedt11,Hultgren12,eshetu14,LinstedtO15} and anchor modeling \cite{RegardtRBJW09,RonnbackRBJW10}. Both are actually very close \cite{anchorVSvault13}, with a somewhat easier evolutivity and a non-destructive schema evolution for anchor modeling, but a larger number of objects to manage because of a sixth normal form (6NF) model, as well as non-automated maintenance procedures for timestamps. Data vaults are closer to traditional multidimensional models (they adopt the third normal form -- 3NF) and are supported by a greater number of tools, which made us select data vaults in this work. However, this choice is not definitively settled.

Data vaults were invented to meet industrial needs \cite{Linstedt11}. 
A data vault is defined at the relational, logical level as a linked set of normalized, detail-oriented and history-tracking tables that support one or more functional domains in an organization. It is a hybrid approach encompassing 3NF and the star schema~\cite{Linstedt11,Hultgren12,eshetu14,LinstedtO15}.
 
This architecture allows incremental low-cost schema construction (making easy to modify business rules), seamless integration of unstructured data and loading in near real time. Data vaults also allow petabyte-scale big data management \cite{LinstedtO15}. Finally, artificial intelligence and data mining methods can easily and directly be applied to a data vault \cite{Linstedt15}.

More concretely, a data vault consists of the following main elements \cite{Linstedt11,Hultgren12,eshetu14,LinstedtO15}.
\begin{itemize}
	\item A \emph{hub} is a basic entity that represents a business concept (a dimension hierarchy level in a classical data warehouse, e.g., customer or product). A hub mainly contains a business key.
	\item A \emph{link} materializes an association between two or more hubs. It would correspond to a fact entity in a classical data warehouse.
	\item A \emph{satellite} contains the various descriptive data related to a hub or link.
\end{itemize}

Finally, note that, although they are defined at the logical level by~\cite{Linstedt11}, these concepts can easily be transposed to the conceptual~\cite{JovanovicB12} and physical \cite{KrnetaJM14} levels. An example of conceptual data vault model is shown in Figure~\ref{fig:modelisation}.

\section{Metadata Vault Model} 
\label{modelemeta}

\subsection{Data Corpus} 
\label{corpus}

To illustrate our proposal with a use case that can serve as a proof of concept, we exploit the data corpus constituted by the TECTONIQ project, which aims to highlight the textile industrial heritage of the Lille Metropolis in northern France \cite{KergosienJSC15,tectoniq}. 

This corpus gathers heterogeneous data, supplied by different sources (Table~\ref{tab:corpus}) and must allow various types of analyses. Thence, the corpus is stored in a data lake whose metadata are multidimensionally modeled \cite{Pathirana15}. We propose in this paper to make these metadata evolutive when new data sources pop up.

\begin{table*}[hbt]
	\centering
	\caption{TECTONIQ corpus features}
	\label{tab:corpus}	
	\begin{tabularx}{\textwidth}{@{}l *4{>{\centering\arraybackslash}X}@{}}
    	\hline
		\textbf{Source} & \textbf{Inventory} & \textbf{La Voix du Nord}  \\ \hline
        \textbf{Description} & Industrial building descriptions & Press articles related to the textile industry \\ \hline
        \textbf{Number of documents} & 49 & 1  \\ \hline
        \textbf{Number of instances} & 49 & 30  \\ \hline
        \textbf{Overall size} & 120 KB & 92 KB  \\ \hline
        \textbf{Format} & XML (data) & XML (documents) \\
		\hline                          
	\end{tabularx}
	\begin{tabularx}{\textwidth}{@{}l *4{>{\centering\arraybackslash}X}@{}}
    	\hline
		\textbf{Source} & \textbf{IHRIS Pictures} & \textbf{Books} \\ \hline
        \textbf{Description}  & Photos and plans of buildings and monuments related to the textile industry & Public domain French history books  \\ \hline
        \textbf{Number of documents}  & 30 & 165 \\ \hline
        \textbf{Number of instances}  & 30 & 165 \\ \hline
        \textbf{Overall size}  & 2.78 MB & 1011.25 MB \\ \hline
        \textbf{Format}  & JPEG & PDF \\
		\hline                          
	\end{tabularx}
\end{table*}

\subsection{Conceptual Model} 
\label{modeleconc}

Although the multidimensional metadata model by \cite{Pathirana15} is too large to be reproduced here, let us describe it briefly. It is designed around a set of dimensions with hierarchies: Date/Epoch, Keyword, Location/District/City, Source, Reference/Author; on which ``fact'' entities (without measures) that correspond to the data sources from Table~\ref{tab:corpus} plug in. 

\begin{figure}[hbt]
  \centering
  \includegraphics[width=12.5cm]{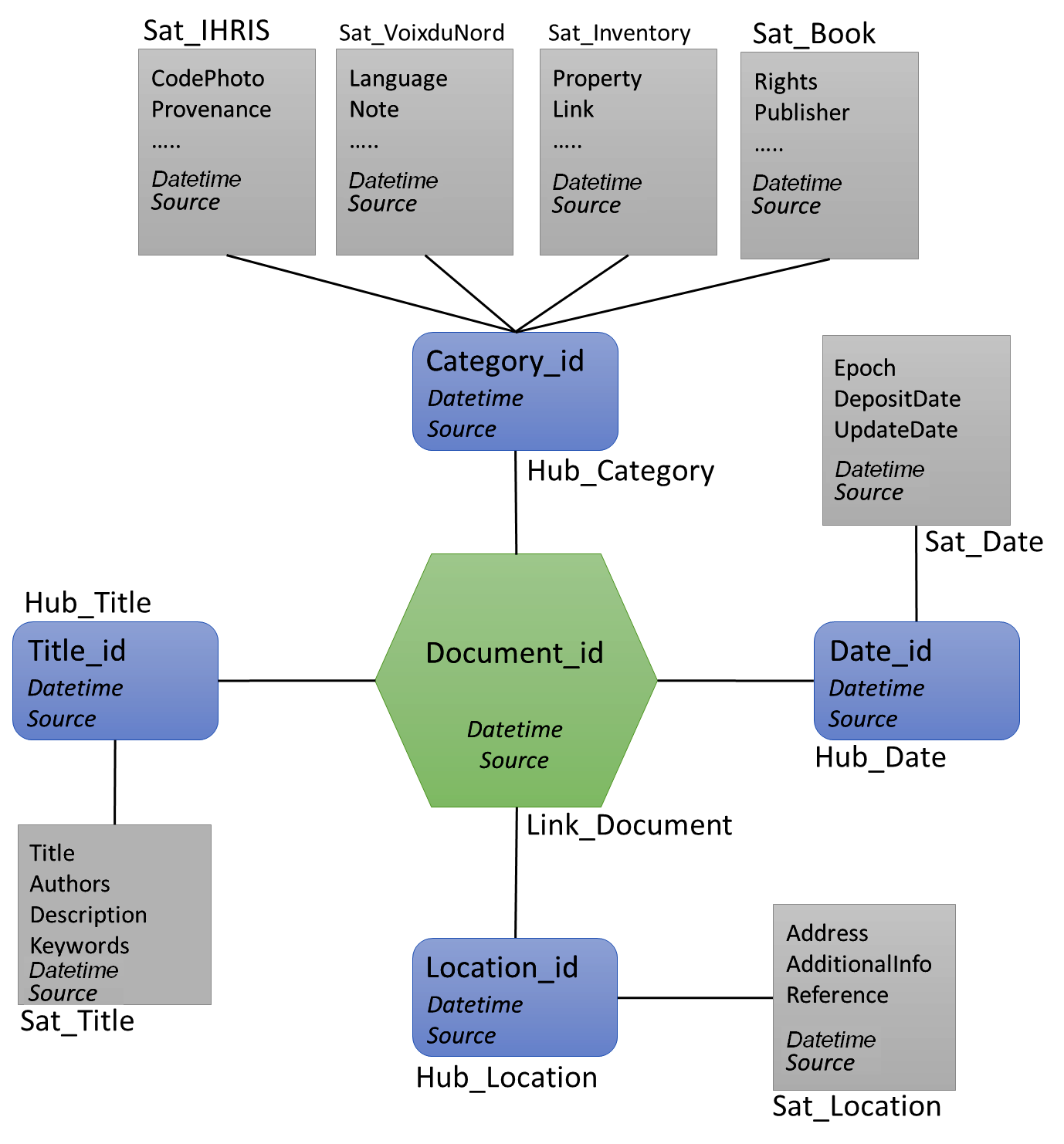}
  \caption{\label{fig:modelisation} Metadata vault conceptual model}
\end{figure}

Figure~\ref{fig:modelisation} illustrates our conceptual metadata vault model, which exploits the conceptual modeling of data vaults' logical, relational concepts \cite{JovanovicB12}. Blue rounded rectangles represent hubs, gray rectangles satellites and the green hexagon a link. Associations between hubs and links are of cardinality ``many to many'' to ensure the greatest generality, while associations betweens hubs or links and satellites are of cardinality ``one to many''.

To identify hubs, which are the most important and most used entities in queries, we target the metadata related to the characteristics of each input document, which are most susceptible to constitute keys. Thus, we select title (\textit{Hub\_Title}), location (\textit{Hub\_Location}), date (\textit{Hub\_Date}) and document category (\textit{Hub\_Category}).

Then, we associate with each of these hubs a satellite that contains the hub's descriptive attributes. For example, the satellite connected to \textit{Hub\_Title}, \textit{Sat\_Title}, contains the \textit{Title}, \emph{Authors}, \emph{Description} and \emph{Keywords} attributes. 

\textit{Hub\_Category} is associated with four satellites that form a classification corresponding to each of the data sources we have. The descriptive attributes of these satellites are specific to each source.

\textit{Link\_Document} allows to associate all hubs. 

Finally, since our model concerns only metadata, each entity (hub, link, satellite) is described by a direct reference (\textit{Source}) to a document (physical file) in the data lake. All data access are made through this reference.

Thanks to ensemble modeling, any new data source or schema evolution within the data corpus can be supported by adding satellites, in the simplest cases, or even hubs and links. In addition, entities that would become obsolete are simply identified by their timestamp (\emph{Datetime}).

\subsection{Logical and Physical Models} 
\label{modeleslogphy}

To show that our conceptual model can adapt to different contexts, and to compare them, we translate it into two types of logical and physical models: a relational model with PostgreSQL and a document-oriented NoSQL model with MongoDB.

The relational metadata model (Table~\ref{Table2})\footnote{We adopt this unusual table presentation to have a uniform representation with the document-oriented model (Table~\ref{Table3}) and facilitate the comparison of the two logical models.} is classically translated from the conceptual model (Figure~\ref{fig:modelisation}). Let us only note that, in Table~\ref{Table2}, the elements of the attributes column that are postfixed by \emph{\_id} are foreign keys, and that satellite primary keys are also foreign keys that reference the corresponding hub's primary key. Eventually, the \emph{Datetime} and \emph{Source} attributes featured in all tables are not repeated for clarity.

\begin{table}[hbt]
\centering
\caption{Metadata vault relational logical model}
\label{Table2}
\begin{tabular}{|l|l|l|}
\hline
\textbf{Table} & \textbf{Primary key} & \textbf{Attributes}                                                                              \\ \hline
\multicolumn{3}{|c|}{\textit{Hubs}}                                                                                        \\ \hline
Hub\_Title      & Title\_id      &                                                                                              \\ \hline
Hub\_Date       & Date\_id       &                                                                                             \\ \hline
Hub\_Location   & Location\_id   &                                                                                              \\ \hline
Hub\_Category   & Category\_id   &                                                                                              \\ \hline
\multicolumn{3}{|c|}{\textit{Link}}                                                                                        \\ \hline
Link\_Document   & Document\_id        & Title\_id, Location\_id, Date\_id, Category\_id \\ \hline
\multicolumn{3}{|c|}{\textit{Satellites}}                                                                                  \\ \hline
Sat\_Title     &  Title\_id,  Datetime             & Title, Authors,  Description, Keywords                                                                                    \\ \hline
Sat\_Date      &  Date\_id,  Datetime             & Epoch, DepositDate,  UpdateDate                                                                                     \\ \hline
Sat\_Location  &  Location\_id,  Datetime& Address, AdditionalInfo, Reference                                                                                  \\ \hline
Sat\_IRHIS      &  Category\_id,  Datetime          & CodePhoto, Provenance                                                                                 \\ \hline
Sat\_VoixDuNord &  Category\_id,  Datetime          & Language, Note                                                                                   \\ \hline
Sat\_Inventory &  Category\_id,  Datetime           & Property, Link                                                                                   \\ \hline
Sat\_Book      &  Category\_id,  Datetime           & Rights, Publisher                                                                                  \\ \hline
\end{tabular}
\end{table}

\begin{table}[ht!]
\centering
\caption{Metadata vault document-oriented logical model}
\label{Table3}
\begin{tabular}{|l|l|l|}
\hline
\textbf{Collection} & \textbf{ObjectID}     & \textbf{Fields}                                                                                         \\ \hline
\multicolumn{3}{|c|}{\textit{Hubs}}                                                                                        \\ \hline
Hub\_Title      & Title\_id    &                                                                                \\ \hline
Hub\_Date       & Date\_id     &                                                                                \\ \hline
Hub\_Location   & Location\_id &                                                                                \\ \hline
Hub\_Category   & Category\_id &                                                                                \\ \hline
\multicolumn{3}{|c|}{\textit{Link}}                                                                                        \\ \hline
Link\_Document   & Document\_id      &  \\ \hline
\multicolumn{3}{|c|}{\textit{Satellites}}                                                                                  \\ \hline
Sat\_Title     &  Title\_id              & Title, Authors, Description, Keywords                                                                                    \\ \hline
Sat\_Date      &  Date\_id               & Epoch, DepositDate, UpdateDate                                                                                     \\ \hline
Sat\_Location  &  Location\_id           & Address, AdditionalInfo, Reference                                                                                  \\ \hline
Sat\_IRHIS      &  Category\_id          & CodePhoto, Provenance                                                                                 \\ \hline
Sat\_VoixDuNord &  Category\_id          & Language, Note                                                                                 \\ \hline
Sat\_Inventory &  Category\_id           & Property, Link                                                                                 \\ \hline
Sat\_Book      &  Category\_id           & Rights, Publisher                                                                          \\ \hline
\end{tabular}
\end{table}

The  document-oriented metadata model  (Table~\ref{Table3}) exploits the notion of collection. Each hub, link and satellite is translated into a collection comprising various documents, each having an identifier (\emph{ObjectID}). In the MongoDB physical model, this identifier is generated automatically when the document is created. As in the relational case, the \emph{Datetime} and \emph{Source} fields present in all collections are not repeated for clarity.

Each hub is associated with both \emph{Link\_Document}  and its satellite(s) through its \emph{ObjectID}. For example, \emph{Hub\_Title} is associated with \emph{Link\_Document} and \emph{Sat\_Title} through \emph{Title\_id}. Finally, Figure~\ref{phymodel-book} presents an example of translation from the document-oriented logical model into the MongoDB physical model in JavaScript Object Notation (JSON) format.

\begin{figure}[hbt]
\centering
\begin{verbatim}{
 "Sat_Book": {
  "00000C842_001": {
   "dc:relation": "http://www.sudoc.fr/122661389",
   "dc:creator": [
    "Petit, Jules",
    "Chambre de commerce et d'industrie (Boulogne-sur-Mer, Pas-de-Calais).",
    "Commission du projet de chemin de fer direct de Calais a Marseille"
   ],
   "dc:subject": "Calais-Marseille, Chemin de fer (France).",
   "dc:publisher": "Villeneuve d'Ascq : SCD Lille 3",
   "dc:date": "[2008]",
   "dc:format": "application/pdf",
   "dc:language": "fre",
   "dc:rights": "domaine public",
   "dc:title": "Projet de chemin de fer de Calais a Marseille rapport 
    fait a la Chambre de commerce de Boulogne, le 25 novembre 1881",
   "dc:type": "text",
   "dc:source": "Bibliotheque Georges Lefebvre"
  }
 }
}\end{verbatim}
\caption{JSON excerpt from the \emph{Sat\_Book} collection of the MongoDB physical model}
\label{phymodel-book}
\end{figure}

At the physical level, metadata are extracted from source documents through a script-based ETL process.

\section{Experimental Validation} 
\label{expes}

The objective of this section is to show the feasibility of our metadata vault model and to compare the physical models of Section~\ref{modeleslogphy} in terms of storage space and query response time.

We conduct our experiments on an Intel Core i5-5300U 2.30 GHz PC with 8~GB of memory, running Windows 64-bit. The DBMSs we use are PostgreSQL 9.6.2-1 and MongoDB ssl 3.4.3. The data corpus in this test set is the one presented in Table~\ref{tab:corpus}.

\subsection{Storage Volume} 
\label{xpstockage}

After inserting metadata in the chosen DBMSs' native format, we measure the storage volume required by PostgreSQL and MongoDB (Table~\ref{table1}). Our measurements take into account indexes automatically created by the DBMSs (e.g., on primary keys in PostgreSQL; no other index is created), unused memory space, and the space that is released when deleting or moving data.

\begin{table}[hbt]
  \centering
  \caption{Metadata volume (KB)}
  \label{table1}
  \begin{tabular}{lrr}
  & \textbf{PostgreSQL} & \textbf{MongoDB} \\
  \hline
    Hub\_Title & 80 & 45 \\
    Hub\_Date & 64  & 36  \\
    Hub\_Location & 72  & 49  \\
    Hub\_Category & 64 & 36 \\
  \hline 
    Link\_Document &  48  & 36  \\
  \hline 
    Sat\_Title & 168  & 69  \\
    Sat\_Date & 48  & 36  \\
    Sat\_Location & 56  & 49  \\ 
    Sat\_IHRIS & 16  & 16  \\ 
    Sat\_VoixDuNord & 16 & 7  \\
    Sat\_Inventory & 88  & 45  \\    
    Sat\_Book & 56  & 37  \\  
  \hline \textbf{Total} & \textbf{776} & \textbf{461} \\
 \end{tabular}
\end{table}

We first note that the volume of metadata generated for 245 files is small (less than 1~MB) in both cases. In addition, MongoDB systematically requires less space than PostgreSQL. This is explained by the use of the JSON format, which is very light, in MongoDB. In contrast, in PostgreSQL, each table is stored as a page vector of predetermined size (8~KB), resulting in pages that are not fully filled.

\subsection{Response Time} 
\label{xpreponse}

To measure the performance of the proposed metadata model, we formulate five queries of increasing complexity with respect to the number of hubs involved (and, therefore, to joins via \emph{Link\_Document}), which we apply on both physical models.

\begin{enumerate}

\item Retrieve all documents whose title contains the word ``factory'' (data sources: \emph{Hub\_Title}, \emph{Sat\_Title}).

\item Retrieve all documents whose title contains the word ``factory'' and whose address is ``Tourcoing'' (data sources: \emph{Hub\_ Title}, \emph{Sat\_Title}, \emph{Hub\_Location}, \emph{Sat\_Location}, \emph{Link\_Doc-ument}).

\item Retrieve all documents whose title contains the word ``factory'', whose address is ``Tourcoing'' and whose deposit date is in 2010 (data sources: \emph{Hub\_Title}, \emph{Sat\_Title}, \emph{Hub\_Location}, \emph{Sat\_Location}, \emph{Hub\_Date}, \emph{Sat\_Date}, \emph{Link\_Document}).

\item Retrieve all documents whose title contains the word ``factory'', whose address is ``Tourcoing'', whose deposit date is in 2010 and that belong to the ``book'' category (data sources: \emph{Hub\_Title}, \emph{Sat\_Title}, \emph{Hub\_Location}, \emph{Sat\_Location}, \emph{Hub\_Date}, \emph{Sat\_Date}, \emph{Hub\_Category}, \emph{Sat\_Book}, \emph{Link\_Document}).

\item Retrieve all titles containing the word ``factory'', and then the corresponding documents, whatever their category (data sources: \emph{Hub\_Title}, \emph{Sat\_Title}, \emph{Hub\_Category}, \emph{Sat\_IHRIS} and/or \emph{Sat\_VoixDuNord} and/or \emph{Sat\_Inventory} and/or \emph{Sat\_Book}, \emph{Link\_ Document}).

\end{enumerate}

The translation of all queries in SQL and MongoDB's query language are provided in Appendices~\ref{q:sql} and \ref{q:mongo}, respectively.

\begin{figure}[hbt]
  \centering
  \includegraphics[width=15cm]{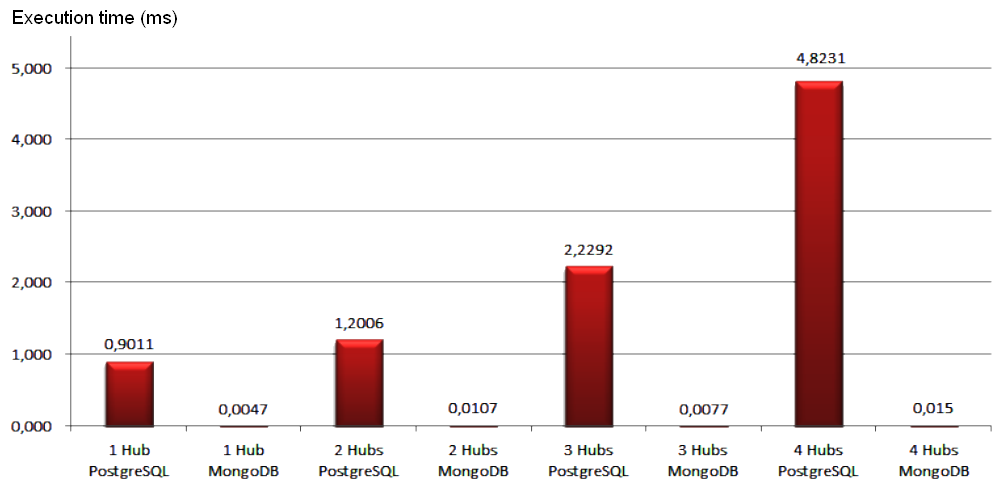}
  \caption{\label{fig:select} Average response times of queries~\#1 to \#4}
\end{figure}

Figure~\ref{fig:select} features the average execution time of queries~\#1 to \#4 under PostgreSQL and MongoDB, respectively. We execute each query 100 times (in the order indicated above) to compensate for any variation in execution time. Figure~\ref{fig:select} shows tremendously low response times for MongoDB, while those obtained with PostgreSQL seem to evolve exponentially.

\begin{figure}[ht!]
  \centering
  \includegraphics[width=10cm]{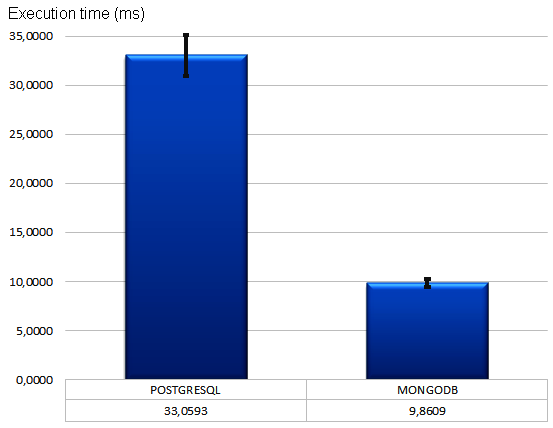}
  \caption{\label{fig:version2} Average response time of query \#5}
\end{figure}

Figure~\ref{fig:version2} illustrates the average execution time of query~\#5, still for 100 executions of the query. It is distinct from Figure~\ref{fig:select} because query~\#5 is more complex than the previous four. It is indeed not possible to dynamically determine the category satellite to use. It is therefore necessary to execute query~\#5 in two steps (subqueries): the first to determine the category and the second to retrieve the remaining information knowing the category. These two operations induce much longer response times than those of queries~\#1 to \#4.

MongoDB is once again more efficient (three times faster on average) than PostgreSQL. This difference in query execution time can be explained by the internal query rewriting done by PostgreSQL, which creates a subquery inducing an expensive join~\cite{postgresqlPerf}, while MongoDB uses a search method that allows the rapid union of collections.

\section{Conclusion} 
\label{conclusion}

From the modeling of data lake metadata as a multidimensional schema, noticing that schema evolution is not guaranteed, we instead propose in this paper an ensemble model, more precisely a metadata vault.

The translation of our conceptual metadata model into logical and physical models, as well as experiments carried out on the TECTONIQ corpus, help demonstrate the feasibility of our modeling choice in terms of metadata storage volume and response time to queries formulated on the metadata.

The comparison of two physical models (PostgreSQL and MongoDB) also reveals the superiority of document-based models for storing this type of metadata.

The perspectives that this preliminary work is opening are numerous. In particular, it would be necessary to test the robustness of the metadata model when source data scale up, add data sources to verify the relevance of data vault modeling and test the model with queries more complex than projections/restrictions.

Moreover, although the supporters of anchor modeling argue that 6NF modeling allows good response times thanks to the join eliminating technique used in modern optimizers \cite{paulley00}, it is also important to compare query efficiency over data vault and anchor models vs. classical star-like schemas, with a benchmark that allows significant database scale-up.

Finally, various alternative metadata models, independently from the modeling techniques used, could be considered and compared. Schemas of the TECTONIQ documents could also be automatically extracted to enrich the metadata.

\section*{Acknowledgments}

The authors would like to thank \'{E}ric Kergosien, leader of the TECTONIQ project, for making the data corpus available.

\bibliographystyle{abbrv}
\bibliography{vaultlake}

\section*{Appendices}
\appendix

\section{SQL Queries} 
\label{q:sql}
\begin{verbatim}
-- 1
SELECT * FROM Hub_Title HT, Sat_Title ST
WHERE HT.Title_id = ST.Title_id
AND Title LIKE '%factory%'

-- 2
SELECT * FROM Link_Document LD, Hub_Title HT, Sat_Title ST, 
   Hub_Location HL, Sat_Location SL
WHERE LD.Title_id = HT.Title_id AND LD.Location_id = HL.Location_id
AND HT.Title_id = ST.Title_id AND HL.Location_id = SL.Location_id
AND Title LIKE '%factory%' AND Address = 'Tourcoing'

-- 3				
SELECT * FROM Link_Document LD, Hub_Title HT, Sat_Title ST, 
   Hub_Location HL, Sat_Location SL, Hub_Date HD, Sat_Date SD
WHERE LD.Title_id = HT.Title_id AND LD.Location_id = HL.Location_id 
AND LD.Date_id = HD.Date_id AND HT.Title_id = ST.Title_id 
AND HL.Location_id = SL.Location_id AND HD.Date_id = SD.Date_id
AND Title LIKE '%factory%' AND Address = 'Tourcoing' 
AND DATE_TRUNC('year', DepositDate) = 2010

-- 4 			
SELECT * FROM Link_Document LD, Hub_Title HT, Sat_Title ST, 
   Hub_Location HL, Sat_Location SL, Hub_Date HD, Sat_Date SD, 
   Hub_Category HC, Sat_Book SB
WHERE LD.Title_id = HT.Title_id AND LD.Location_id = HL.Location_id 
AND LD.Date_id = HD.Date_id AND LD.Category_id = HC.Category_id
AND HT.Title_id = ST.Title_id AND HL.Location_id = SL.Location_id 
AND HD.Date_id = SD.Date_id AND HC.Category_id = SB.Category_id
AND Title LIKE '%factory%' AND Address = 'Tourcoing' 
AND DATE_TRUNC('year', DepositDate) = 2010 AND SB.Category_id = 'book'

-- 5 
-- PL/pgSQL is needed here.
DECLARE
   doc RECORD;
   cat RECORD;
BEGIN
   FOR doc IN 
      SELECT * FROM Link_Document LD, Hub_Title HT, Sat_Title ST
      WHERE LD.Title_id = HT.Title_id AND	HT.Title_id = ST.Title_id
      AND Title LIKE '%factory%
   LOOP
      FOR cat IN EXECUTE
         "SELECT * FROM Hub_Category HC, Sat_" || doc.category_id || " SL
        WHERE HC.Category_id = SL.Category_id"
      LOOP
         -- Tuples doc and cat together provide full document info.
      END LOOP;
   END LOOP;
END;
\end{verbatim}

\section{MongoDB Queries} 
\label{q:mongo}
\begin{verbatim}
# 1
db.hub_title.aggregate([ 
        { $match : { 'title': /factory/  }},
        { $lookup : {
                from : "sat_title",
                localField : "id",
                foreignField : "id",
                as : "satellite" }}
]);

# 2
db.hub_title.aggregate([ 
        { $match : { 'title': /factory/  }},
        { $lookup : {
                from : "sat_title",
                localField : "id",
                foreignField : "id",
                as : "satellite" }},
		{ $lookup : {
                from : "link_document",
                localField : "id",
                foreignField : "id",
                as : "hub_location" }},
        {$unwind : "$link_document"},
        { $lookup : {
                from : "hub_location",
                localField : "id",
                foreignField : "id",
                as : "hub_location" }},
        { $lookup : {
                from : "sat_location",
                localField : "id",
                foreignField : "id",
                as : "sat_location" }},
        {$unwind : "$hub_location"},
        {$match : {'hub_location.location': /Tourcoing/}}
]);

# 3				
db.hub_title.aggregate([ 
        { $match : { 'title': /factory/  }},
        { $lookup : {
                from : "sat_title",
                localField : "id",
                foreignField : "id",
                as : "satellite" }},
		{ $lookup : {
                from : "link_document",
                localField : "id",
                foreignField : "id",
                as : "hub_location" }},
        {$unwind : "$link_document"},
        { $lookup : {
                from : "hub_location",
                localField : "id",
                foreignField : "id",
                as : "hub_location" }},
        { $lookup : {
                from : "sat_location",
                localField : "id",
                foreignField : "id",
                as : "sat_location" }},
        {$unwind : "$hub_location"},
        {$match : {'hub_location.location': /Tourcoing/}},
		{$lookup : {
                from : "hub_date",
                localField : "id",
                foreignField : "id",
                as : "hub_date" }},
        {$lookup : {
                from : "sat_date",
                localField : "id",
                foreignField : "id",
                as : "sat_date" }},
        {$unwind : "$hub_date"},
        {$match : {'hub_date.depositdate': /2010/}}
]);

# 4				
db.hub_title.aggregate([ 
      { $match : { 'title': /factory/  }},
      { $lookup : {
               from : "sat_title",
               localField : "id",
               foreignField : "id",
               as : "satellite" }},
      { $lookup : {
               from : "link_document",
               localField : "id",
               foreignField : "id",
               as : "hub_location" }},
      { $unwind : "$link_document"},
      { $lookup : {
               from : "hub_location",
               localField : "id",
               foreignField : "id",
               as : "hub_location" }},
      { $lookup : {
               from : "sat_location",
               localField : "id",
               foreignField : "id",
               as : "sat_location" }},
      {$unwind : "$hub_location"},
      {$match : {'hub_location.location': /Tourcoing/}},
      {$lookup : {
               from : "hub_date",
               localField : "id",
               foreignField : "id",
               as : "hub_date" }},
      {$lookup : {
               from : "sat_date",
               localField : "id",
               foreignField : "id",
               as : "sat_date" }},
      {$unwind : "$hub_date"},
      {$match : {'hub_date.depositdate': /2010/}},
      {$lookup : {
               from : "hub_category",
               localField : "id",
               foreignField : "id",
               as : "hub_category" }},
      {$unwind : "$hub_category"},
      {$match : {'hub_category.category': /book/}},
      {$lookup : {
               from : "sat_books",
               localField : "id",
               foreignField : "id",
               as : "sat_books" }}
]);

# 5
# Two steps and a function are needed here.
var docs = db.hub_title.aggregate([ 
      { $match : { 'title': /factory/  }},
      { $lookup : {
               from : "sat_title",
               localField : "id",
               foreignField : "id",
               as : "satellite" }},
      { $lookup : {
               from : "link_document",
               localField : "id",
               foreignField : "id",
               as : "link_document" }},
      {$unwind : "$link_document"},
      {$project : { "cat" : "$link_document.category_id" }
]);
docs.forEach(function(cat){ 
   db.hub_category.aggregate([ 
      {$lookup : {
               from : "sat_$cat",
               localField : "id",
               foreignField : "id",
               as : "cat" }},
      {$unwind : "cat"}
   ]);
});
\end{verbatim}

\end{document}